\documentclass[12pt]{article}
\usepackage{graphics}
\begin{document}
\begin{center}
{\normalsize Effect of the Constant Potential in the Bethe-Salpeter Equation \\ for Proton-Neutron Elastic Scattering}
\end{center}
\begin{center}
{Susumu Kinpara}
\end{center}
\begin{center}
{\it National Institute of Radiological Sciences \\ Chiba 263-8555, Japan}
\end{center}
\begin{abstract}
Nucleon-nucleon potential is given in coordinate space starting from the Bethe-Salpeter equation
appropriate to elastic scattering at the intermediate energy region.
The next order in the potential is added to construct the scattering wave for ${}^3S_1$ state.
The constant terms have an effect on the spin observables particularly at the center of mass angle $\theta=90^{\circ}$.
\end{abstract}
\hspace*{4.mm}
Nucleon-nucleon interaction is significant to draw conclusions on the nuclear system.
Particularly the elastic scattering experiments provide many observables as the incident energy increases and
which are useful to study the fundamental properties of the nuclear force from the pion tail to the short-range region.
For them the meson-exchange model with the pseudovector coupling pion interaction is available.
Then the off-shell behavior should be treated by the Bethe-Salpeter (BS) equation 
of the spinor-spinor type $\cite{Seto}$.
\\\hspace*{4.mm}
The non-perturbative character is reflected upon the potential 
and for the calculation of the nuclear system it is taken into account as much as possible.
One of the interesting properties of the BS equation is the relative time 
and the effects of the retardation in the propagators and the vertex part of pion-nucleon coupling are indispensable
to treat the potential part exactly.
To obtain the observables on two-nucleon system the region in which the relative time $t$ = 0 is sufficient 
within the framework of the relativity.
\\\hspace*{4.mm}
In our previous study the BS equation has been expanded by a set of the $\Gamma$ matrices 
and which results in the simultaneous equations $\cite{Kinpara}$.
By virtue of the auxiliary relation the first derivatives of the components with respect to $t$ at $t$ = 0 are permitted to drop and the equations become approximately the equivalent form with the Schr$\ddot{\rm o}$dinger equation.
Therefore various ways to solve the equation under the singular potential $\cite{Case}$ are applicable 
for the investigation of the elastic scattering. 
\\\hspace*{4.mm}
To proceed calculations the equations in momentum space are converted to the ones in coordinate space
by the Fourier transform. 
They are given as follows
\\\begin{eqnarray}
(-k^2/M+\partial^\mu\partial_\mu/M)\psi_P(x)+M^{-1}P_0\,F[\psi_i, \psi_P,\psi_a]=-V_P(x)\psi_P(x),
\end{eqnarray}
\begin{eqnarray}
&&(k^2/M-\partial^\mu\partial_\mu/M)\psi_a(x)+2M^{-1}\,\partial_0^2\,\psi_a(x)+2\,F[\psi_i, \psi_P,\psi_a]\nonumber\\
&&=
V_a(x)\psi_a(x)
+V_a^i(x)\psi_i(x),
\end{eqnarray}
\begin{eqnarray}
&&(-k^2/M+\partial^\mu\partial_\mu/M)\psi_i(x)-2M^{-1}\,\partial_0^2\,\psi_i(x)+iM^{-1}\partial_i\,F[\psi_i, \psi_P,\psi_a]\nonumber\\
&&=
-V_i^j(x)\psi_j(x)
-V_i^a(x)\psi_a(x),
\end{eqnarray}
\begin{eqnarray}
F[\psi_i, \psi_P,\psi_a]\equiv 2\,i\,\partial^i\,\psi_i(x)+M\,\psi_a(x)+\frac{P_0}{2}\,\psi_P(x),
\end{eqnarray}\\
in which both sides of the equations are divided by the nucleon mass $M$ so that
the connection to the Schr$\ddot{\rm o}$dinger equation is made clear.
Here $\it\psi_P(x)$, $\it\psi_a(x)$ and $\it\psi_i(x)$ ($i=1,2,3$) denote
the pseudoscalar, the axial-vector and the $i$-th component of the tensor waves respectively. 
\\\hspace*{4.mm}
In the center of mass system $k \equiv \sqrt{P_0^2/4-M^2}$ 
and it depends on the total energy $P_0$ of the two-body system.
When we apply the equations to elastic scattering between two nucleons
the kinetic energy $T$ of the incident nucleon in the laboratory system is needed as the input parameter.
Using the relation of the Lorentz transformation between the laboratory system and the center of mass system
it is shown that $T$ and $k$ are connected by the relation $k = \sqrt{M T / 2}$.
\\\hspace*{4.mm}
To deduce the observables on two-nucleon system 
the dependence of the solutions on relative time $t$ is neglected by restricting
the region of space-time at $t=0$.
The instantaneous approximation about the propagation of meson corresponds to the nonrelativistic treatment applicable
to the phenomena.
Then the correction of the terms $\sim\partial^2/\partial t^2$ in Eqs. (1) $\sim$ (3) may supply 
the alternative way to treat the relativistic effect.
\\\hspace*{4.mm}
Each of three simultaneous equations (Eqs. (1) $\sim$ (3)) is connected with one another 
by the term $F[\psi_i,\psi_P,\psi_a]$.
It plays an important role to investigate the quadrupole moment of deuteron.
The tensor equation (Eq. (3)) enables us to describe the spin triplet state.
The contribution of three waves may cancel to a degree
so as to make the value of $F[\psi_i, \psi_P,\psi_a]$ nearly equal to zero in the asymptotic approximation.
So in order to do computations for elastic scattering easily we follow the speculation and drop the terms
on $F[\psi_i, \psi_P,\psi_a]$ breaking the usual spin-orbit structure.
\\\hspace*{4.mm}
Both the pseudoscalar equation (Eq. (1)) and the axial-vector equation (Eq. (2)) describe the spin singlet 
and then the tensor equation (Eq. (3)) corresponds to the spin triplet states.
The dynamical properties of two-nucleon system are governed by the potential 
\\\begin{eqnarray}
V_P(x)\equiv iM^{-1}\sum_{i\,=\,\sigma,\omega,\pi,\rho}\lambda_i^P\,G_i\,v(x)_i,
\end{eqnarray}
\begin{eqnarray}
V_a(x)\equiv iM^{-1}\sum_{i\,=\,\sigma,\omega,\pi,\rho}(-\lambda_i^A)\,G_i \,v(x)_i\qquad\nonumber\\
+2iM^{-1}\frac{G_\pi}{m_\pi^2}v_1(x)_\pi
+4iM^{-1}G_\rho a^2 v_1(x)_\rho,
\end{eqnarray}
\begin{eqnarray}
V_j^k(x)\equiv iM^{-1}\sum_{i\,=\,\sigma,\omega,\pi,\rho}\lambda_i^T\,G_i \,v(x)_i\,\delta_j^k\qquad\qquad\nonumber\\
-4iM^{-1}\frac{G_\pi}{m_\pi^2}v_1(x)_\pi\,\delta_j^k-2iM^{-1}\frac{G_\pi}{m_\pi^2}\frac{x_j x^k}{r^2}v_2(x)_\pi,
\end{eqnarray}
\begin{eqnarray}
V_a^i(x)\equiv -4M^{-1}\,G_\rho\, a\, x^i\, v_1(x)_\rho = -4V^a_i(x),
\end{eqnarray}\\
where 
$G_\sigma = g_\sigma^{\rm 2}$, 
$G_\omega = g_\omega^{\rm 2}$, 
$G_\pi = \vec{\tau}_1\cdot\vec{\tau}_2\,f_\pi^2 $ and
$G_\rho = \vec{\tau}_1\cdot\vec{\tau}_2\,g_\rho^{\rm 2}$
are the strength of the meson-nucleon coupling respectively 
and the values of $\lambda^{P,A,T}_{\sigma,\omega,\pi,\rho}$ are seen in Ref. $\cite{Kinpara}$.
The mixing of the vector-tensor coupling of $\rho$ meson is defined by $a \equiv f_\rho g^{-1}_\rho/2 M$.
We have omitted the terms which become zero at $x_0 = 0$.
\\\hspace*{4.mm}
The spatial dependence of the potential is determined 
by the Feynman propagator $\Delta_F(x-x^\prime,m_\sigma)\equiv -i\,\langle \, T\phi(x)\phi(x^\prime) \, \rangle$ 
and $v(x)_\sigma \equiv -\Delta_F(x,m_\sigma)$ in the case of the $\sigma$ meson exchange interaction.
For the other mesons the notation of the potential is analogous to it.
At present our interest is limited to the relative time $x^0 = 0$ 
and the explicit form of $v(x)$ at the space-like region ($x^2 < 0$) is 
\\\begin{eqnarray}
v(x) = \frac{i}{(2\pi)^2}\cdot\frac{m}{z}\,K_1(m z),\qquad(z\equiv\sqrt{-x^2}>0) 
\end{eqnarray}
with the mass $m$.
In Eq. (9) the subscripts $i$ of $m_i$ and $v(x)_i$ are dropped for simplicity.
The first and the second derivatives of $v(x)$ are 
\begin{eqnarray}
\partial_\mu v(x) = -x_\mu v_1(x),
\end{eqnarray}
\begin{eqnarray}
\partial_\mu\partial_\nu v(x) = -g_{\mu\nu}v_1(x)-\frac{x_\mu x_\nu}{z^2}v_2(x),
\end{eqnarray}
\begin{eqnarray}
v_1(x) \equiv -\frac{i}{(2\pi)^2}\cdot\frac{m^2}{z^2}\,K_2(mz),
\end{eqnarray}
\begin{eqnarray}
v_2(x) \equiv -\frac{i}{(2\pi)^2}\cdot\frac{m^3}{z}\,K_3(mz).
\end{eqnarray}\\
Here $K_i(z)$ ($i=1,2,3$) is the modified Bessel function of the second kind.
\\\hspace*{4.mm}
The pseudoscalar potential ($V_P(x)$) is in the form of
the sum of the boson propagators multiplied by the respective factors above.
At the distance $|\vec{x}|\rightarrow 0$ 
the inverse square part represents the properties of the potential such as $\sim 1/|\vec{x}|^2$.
The computation of the matrix element is thus done without the difficulty of divergence.
\\\hspace*{4.mm}
On the other hand the axial-vector potential ($V_a(x)$) and the tensor potential ($V_j^k(x)$) have
the terms of $v_1(x)$ and $v_2(x)$.
They give the inverse fourth power potential in the leading order at $|\vec{x}|\rightarrow 0$.
To make the computations tractable and furthermore obtain the convergent result in the Born term 
we introduce the regulator for the pion propagator in momentum space
\\\begin{eqnarray}
\frac{\Lambda^2}{\Lambda^2-p^2}\cdot\frac{1}{m^2-p^2-i\epsilon}.
\end{eqnarray}\\
Including the cut-off factor the interactions $v(x)$, $v_1(x)$ and $v_2(x)$ change as
\\\begin{eqnarray}
v(x)\rightarrow\frac{\Lambda^2}{\Lambda^2-m^2}(v(x)-\frac{\Lambda^2}{m^2}v(x)_\Lambda),\qquad\qquad
\end{eqnarray}
\begin{eqnarray}
v_i(x)\rightarrow\frac{\Lambda^2}{\Lambda^2-m^2}(v_i(x)-v_i(x)_\Lambda).\quad(i=1,2)
\end{eqnarray}
The subscripts $\it\Lambda$ in $v(x)_\Lambda$ and $v_i(x)_\Lambda$ stand for the replacement of the mass from $m$ to $\it\Lambda$.
\\\hspace*{4.mm}
Determining the value of the cut-off parameter $\it\Lambda$
the higher component of the four-momentum transfer ($|q| \geq \Lambda$)
or the short-range part of the potential ($|\vec{x}| \leq 1/\Lambda$) is suppressed effectively.
Thus the inclusion of $\it\Lambda$ makes the leading terms of $v_1(x)$ and $v_2(x)$ expanded in powers of $z$ change 
from the original form $\sim z^{-4}$ to $\sim z^{-2}$, appropriate to the computation of the Born term.
The use of $\it\Lambda$ has another meaning that the short-range part of one-pion exchange potential is partly
substituted by the heavier mesons such as $\sigma$ meson expressing the two-pion correlation. 
The tentative value $\Lambda\sim 500$ MeV is adopted here 
taking account of the calculation of the binding energy of deuteron.
\\\hspace*{4.mm}
In our previous study we have used the approximation to the ${}^{3}S_{1}$ state of elastic scattering
by the solution under the inverse square potential part of $V_j^k(x)$.
To determine the form of the ${}^{3}S_{1}$ state more accurately 
the potential $V_j^k(x)$ in the tensor equation (Eq. (7)) is expanded in powers of $r\equiv \vert\vec{x}\vert$ 
and left up to the constant order as 
\\\begin{equation}
V_T(r) = V_{-2}(r) + V_{0}(r) +O(r^2), \qquad\qquad\qquad\qquad\qquad\qquad
\end{equation}
\begin{equation}
V_{-2}(r) = -\frac{1}{(2\pi)^2 M r^2}
(\lambda_\sigma^T G_\sigma
+\lambda_\rho^T G_\rho
+G_\pi \frac{\Lambda^2}{m_\pi^2}(1-2 \, b^J_{LL^\prime})),
\end{equation}
\begin{eqnarray}
V_{0}(r) = -\frac{\lambda_\sigma^T G_\sigma m_\sigma^2}{(4 \pi)^2 M}(-1+2\gamma+2\log\frac{m_\sigma r}{2})
\qquad\qquad\quad\quad\nonumber\\
-\frac{\lambda_\rho^T G_\rho m_\rho^2}{(4 \pi)^2 M}(-1+2\gamma+2\log\frac{m_\rho r}{2})\qquad\qquad\nonumber\\
+\frac{G_\pi\Lambda^2}{2 (4 \pi)^2 M}(1+\frac{\Lambda^2}{m_\pi^2})(1-2 \, b^J_{LL^\prime}),
\end{eqnarray}
\begin{equation}
b^J_{LL^\prime}\equiv (-1)^{\frac{L-L^\prime+2}{2}}\,(1\,0\,J\,0\,\vert\,L\,0)\,(1\,0\,J\,0\,\vert\,L^\prime\,0).
\end{equation}\\
Here $V_T(r)$ denotes the $\langle L \vert V \vert L^\prime \rangle_J$ component 
of the tensor potential $V_j^k(x)$ and in Eq. (19) $\gamma=0.577\cdots$ is the Euler's constant.  
In the present case the value $b^{1}_{00}=-1/3$ is used to obtain the wave function for the ${}^{3}S_{1}$ state.
\\\hspace*{4.mm}
The constant potential $V_0(r)$ contains the $\sim log \, r$ terms 
which give rise to a difference from the simple form
$V_T(r) = -g M^{-1}/r^2 + c_0$.
To incorporate the logarithmic function we substitute it with the approximate form
\begin{equation}
\log \mu r \rightarrow ((\mu r)^\eta -1)/\eta,\quad(\eta=-2)
\end{equation}
giving the coefficients $g$ and $c_0$.
The mass parameter $\mu$ is introduced to make the function be dimensionless.
The value of $\mu$ is determined so as to keep the strength $g=1/4$.
Thus the order $\nu$ of the Bessel function $J_\nu(k r)$ remains at $\nu=\sqrt{(L+1/2)^2-g}\rightarrow0$
from our previous calculation neglecting the $\rho$ meson exchange interaction.
In other words the condition for $g$ is attainable 
provided that the $\sim r^{-2}$ potential arising from the logarithmic functions in $V_0(r)$ 
cancels that of the $\rho$ meson in $V_{-2}(r)$.
Hence $\mu$ is determined as
\\\begin{eqnarray}
\mu_0^2=\frac{m_\rho^2}{4}(1+\frac{m_\sigma^2}{m_\rho^2}
\cdot\frac{\lambda_\sigma^T G_\sigma}{\lambda_\rho^T G_\rho}),
\end{eqnarray}\\
where $\mu_0$ means the specific value of $\mu$.
Using the parameter set of the meson exchange model the value is found to be $\mu_0$ $\sim$ 450 MeV.
\\\hspace*{4.mm}
In order to treat the higher-order terms 
the element of the K-matrix is calculated by means of the exact wave function $\mid \nu>$ \\
\begin{equation}
<L^\prime \mid V(\mit\Lambda_{\rm 0}) \mid \nu>\;=\;<L^\prime \mid V(\mit\Lambda_{\rm 0}) \mid L> F(\mit\Lambda),
\end{equation}
\begin{equation}
F(\mit\Lambda) \equiv F_{\rm 0}(\mit\Lambda) \cdot F_{\rm 1}(\mit\Lambda),
\end{equation}
\begin{equation}
F_0(\mit\Lambda) \equiv \; <L^\prime \mid V(\mit\Lambda) \mid \nu_{\rm 0}>/<L^\prime \mid V(\mit\Lambda) \mid L>,
\end{equation}
\begin{equation}
F_1(\mit\Lambda) \equiv \; <L^\prime \mid V(\mit\Lambda) \mid \nu>/<L^\prime \mid V(\mit\Lambda) \mid \nu_{\rm 0}>.
\end{equation}\\
Here $V(\mit\Lambda_{\rm 0})$ represents each potential by Eqs. (5) $\sim$ (7)
with the cut-off value $\mit\Lambda_{\rm 0}\sim $ 500 MeV in the pion exchange interaction.
$\mid L >$ is the partial wave with the orbital angular momentum $L$.
Since the divergence in the numerator of $F_0(\mit\Lambda)$ cancels out that in the denominator
the limit $\mit\Lambda\rightarrow\infty$ for $V(\mit\Lambda)$ is allowed 
or alternatively the cut-off procedure is not used for $F_0(\mit\Lambda)$ from the outset 
as we have done here ($\,F_{0,1} \equiv F_{0,1}(\infty)\,$). 
In the previous study of proton-neutron elastic scattering $\mid \nu>$ 
for the spin triplet $S$-wave (${}^3 S_1$) has been replaced
by the solution $\mid \nu_0>$ under the inverse square potential accompanying the simplification $F_1$ = 1.
\\\hspace*{4.mm}
Among the spin observables of proton-neutron elastic scattering
it has been seen in the spin correlation parameter $A_{yy}(\theta)$
that the numerical result of the calculation by the $F_1$ = 1 approximation 
is distinct from the experimental data $\cite{Bandyopadhyay}$.
It would be attributed to the form of the wave function for the ${}^3 S_1$ state because the enhancement
of the matrix element about 20 $\%$ done tentatively increases $A_{yy}(\theta)$ at $\theta = 90^\circ$ 
largely toward the experimental data.
Then we need to determine $F_1$ rigorously to study the relation between the potential and the elastic scattering.
\\\hspace*{4.mm}
In order to derive the factor $F_1$ the approximate wave function is used instead of $\mid \nu>$ 
by keeping the leading plus next to leading terms in the potential $V_T(r)$ (Eq. (17)).
Shifting $k$ in the solution under the inverse square potential 
such as $k \rightarrow k^\prime = \alpha k$ (\,$\alpha\equiv\sqrt{1 - 2 c_0/T}$\,)
the constant term is incorporated into the equation conveniently.
It is as follows
\begin{equation}
\psi(r) = \sqrt{\frac{\pi}{2 kr}}(A\,J_\nu(\alpha k r)+B\,N_\nu(\alpha k r)),
\end{equation}
\begin{equation}
A=f({\rm cos}\,\omega+{\rm tan}\delta_L\,{\rm sin}\,\omega),
\end{equation}
\begin{equation}
B=f({\rm sin}\,\omega-{\rm tan}\delta_L\,{\rm cos}\,\omega),
\end{equation}
\begin{equation}
f\equiv\frac{1+\alpha}{\rm 2 \sqrt{\alpha}} \quad(\,\geq 1\,),
\end{equation}\\
where $\omega\equiv\pi(L+1/2-\nu)/2$ and $\delta_L$ is the phase-shift parameter of the elastic scattering
under the potential $V_{-2}(r)+V_{0}(r)$ in Eqs. (17) $\sim$ (20). 
To derive the coefficients $A$ and $B$ 
the boundary condition is used connecting the solution inside the boundary with the outer wave at large distances.
The oscillatory parts stemmed from the constant potential $c_0$ are not included to proceed the calculation.
\\\hspace*{4.mm}
When $c_0 = 0$ ($\alpha = 1$) the phase-shift $\delta_L$ is indefinite 
and to determine $\delta_L$ and the normalization constant $A$ 
the Neumann function part has to be set equal to zero ($B$ = 0).
The Bessel function $J_\nu(\alpha k r)$ is substituted by the leading term
as $J_\nu(\alpha k r) \approx \alpha^\nu J_\nu(k r)$.
Whether the approximation to $J_\nu(\alpha k r)$ is valid or not is dependent upon the average range $\langle kr \rangle$
the interaction works, however, it is not yet verified.
Consequently $F_1$ results in the simple form
$F_1 = \alpha^{\nu}f$, where
$\nu = \sqrt{(L+\frac{1}{2})^2-g} \rightarrow 0 $ and making us set $\nu=0$.
\\\hspace*{4.mm}
$F_1$ is calculated by using the value of the laboratory energy $T$ = 500 MeV. 
The result is $F_1\approx$ 1.05, which is too small 
in comparison with the desirable value $F_1\sim$ 1.2 to reproduce 
the experimental data of $A_{yy}(\theta=90^\circ)$ at the intermediate energy region.
To improve the result the procedure is applied to the $B\neq$ 0 case.
By supposing the condition $\psi(r_0)=0$ for the wave function in Eq. (27) 
at a short-range region $r_0$ the coefficient $B$ is determined as
\\\begin{eqnarray}
B=f\,\frac{J_\nu(\alpha k r_0)}{J_\nu(\alpha k r_0)\,{\rm sin}\,\omega-N_\nu(\alpha k r_0)\,{\rm cos}\,\omega},
\end{eqnarray}
and therefore due to the change of the coefficient $A$ in the Bessel function part the factor $F_1$ shifts to 
\\\begin{eqnarray}
F_1 = \alpha^{\nu}(f-B\,{\rm sin}\,\omega).
\end{eqnarray}
The zero point $r_0$ may represent the radius of the repulsive core between two nucleons 
which is a characteristic of the nuclear force.
It is interpreted that the generation of the mass of pion by the chiral symmetry breaking is
followed by the core in the case of the elastic scattering.
The situation is analogous to the bound state 
in which the residual part of the interaction excluding the inverse square potential is essential to construct deuteron.
\\\hspace*{4.mm}
One of the features of $F_1$ is the dependence on the core radius $r_0$.
Since $B$ is divergent at $r_0=r_b$ giving the relation $J_\nu(\alpha k r_b)=N_\nu(\alpha k r_b)$ in Eq. (31) the value of
$r_0$ is taken to be $r_0>r_b$ to examine the trend of $F_1$ on $r_0$.
Below $r_b$ there is no region of $r_0$ in which the appropriate value $F_1\sim$ 1 is attainable.
When $T$ = 500 MeV the divergent point is at $r_b\sim$ 0.4 fm and it is roughly the same value as the one 
used for the calculation of the binding energy of deuteron.
Fig. 1 shows the factor $F_1$ plotted as a function of $r_0$ at the laboratory energy $T$ = 500 MeV.
It gives the reasonable value $F_1\approx$ 1.2 at the $r_0$ a rather larger than $r_b$ ($r_0\approx$ 1.25\,$r_b$).
\\\hspace*{4.mm}
The effect of the Neumann function part on $F_1$ is interesting and the result arising from it is shown below.
When $\nu=0$ the Neumann function part $N_0(\alpha k r)$ in Eq. (27) is expressed
by means of the Bessel function $J_0(\alpha k r)$ as 
\begin{eqnarray}
N_0(\alpha k r)\sim \frac{2}{\pi}\,(\gamma+{\rm log} (\alpha k r/2))\,J_0(\alpha k r)+O((\alpha k r)^{2}),
\end{eqnarray}
neglecting the remaining terms in the order of $O((\alpha k r)^{2})$.
The ${\it log} (\alpha k r/2)$ is replaced by the constant ${\it log}(\alpha \langle k r \rangle/2)$
using the suitable value $\langle k r \rangle \rightarrow k r_0$.
Thus the final result of $F_1$ yields
\\\begin{eqnarray}
F_1
= \alpha^{\nu}(f-B\,{\rm sin}\,\omega+B\,{\rm cos}\,\omega\,{\rm \frac{2}{\pi}}
(\gamma+{\rm log}\frac{\alpha k r_{\rm 0}}{\rm 2})).
\end{eqnarray}
The way of the logarithmic function is acceptable
because $F_1$ is not sensitive much on the choice of $\langle k r \rangle$ 
such that the 10 $\%$ shift of $\langle k r \rangle$ moves $F_1$ roughly 1 $\%$.
As seen in Fig. 1 inclusion of the Neumann part makes $F_1$ (Eq. (34)) lower 
and in other words which reduces the size of the core radius favorably.
\\\hspace*{4.mm}
While at the $T$ = 500 MeV region the correction of $F_1$ works well 
it does not explain the data of $A_{yy}(\theta=90^\circ)$ at the $T$ = 200 MeV region. 
Since the approximation $F_1 \sim$ 1.0 is appropriate in the region of $T$
the $c_0 = 0$ solution is sufficient to reproduce the experimental data. 
Then the $F_1$ leaves much to be improved on the $T$ dependence.
An additional change of $c_0$ may vary the form of the wave function in Eq. (27) 
so as to provide the most suitable values of the core radius $r_0$.
Without it $r_0$ tends to increase as the energy $T$ decreases for the optimum value of $F_1$.
As seen from $\alpha=\sqrt{1-c_0 M/k^2}$ the dependence of $c_0$ on $k$ is possibly required
to make $F_1$ stay in $1\sim1.2$. 
\\\hspace*{4.mm}
The result of the spin correlation parameter $A_{yy}(\theta)$ is shown in Fig. 2 
as a function of the scattering angle $\theta$ in the center of mass system at the laboratory energy $T$ = 425 MeV 
using three values $F_1$ = 1.1, 1.15 and 1.2. 
As the value of $F_1$ increases the curve rises and intersects the 
experimental value $\cite{Bandyopadhyay}$ at $\theta=90^\circ$ consequently.
Since the present formulation of the density matrix assumes the symmetry of the isospin
under two identical nucleons there exists the relation $A_{yy}(\pi/2-\theta)=A_{yy}(\theta)$
in the result of the calculation.
It appears that the theoretical value of $A_{yy}(\theta=90^\circ)$ is proportional to the size of $F_1$.
A simple form
$A_{yy}(\theta)\sim{\rm Re \,[\,M^{0}_{11}(\theta)\,M^{0}_{1-1}(\theta)\,]} 
\sim F_1\cdot({\rm cos \,2}\theta -{\rm 1})$
may represent the dependence of $A_{yy}(\theta)$ on $\theta$ 
using the isospin singlet part of the M matrix and therefore it has a minimum at $\theta$ = $90^\circ$.
\\\hspace*{4.mm}
In the present study the relative time dependence is neglected to apply to calculations for elastic scattering.
Removing the derivatives on the relative time 
the equations are analogous to the Schr$\ddot{\rm o}$dinger equation and thus the phase-shift analysis method is applicable.
As the energy decreases the core radius is required to enlarge to interpret the experimental data 
and it begins to occupy the short-range region of the nuclear force.
To prevent $r_0$ from moving an effect is expected in addition to the constant potential although it is not included here.
Besides the corrections with the relative time the higher-order that is the square potential 
next to the constant term is feasible.
\hspace*{4.mm}

\newpage
\begin{figure}
\begin{center}
\scalebox{0.5}{\includegraphics{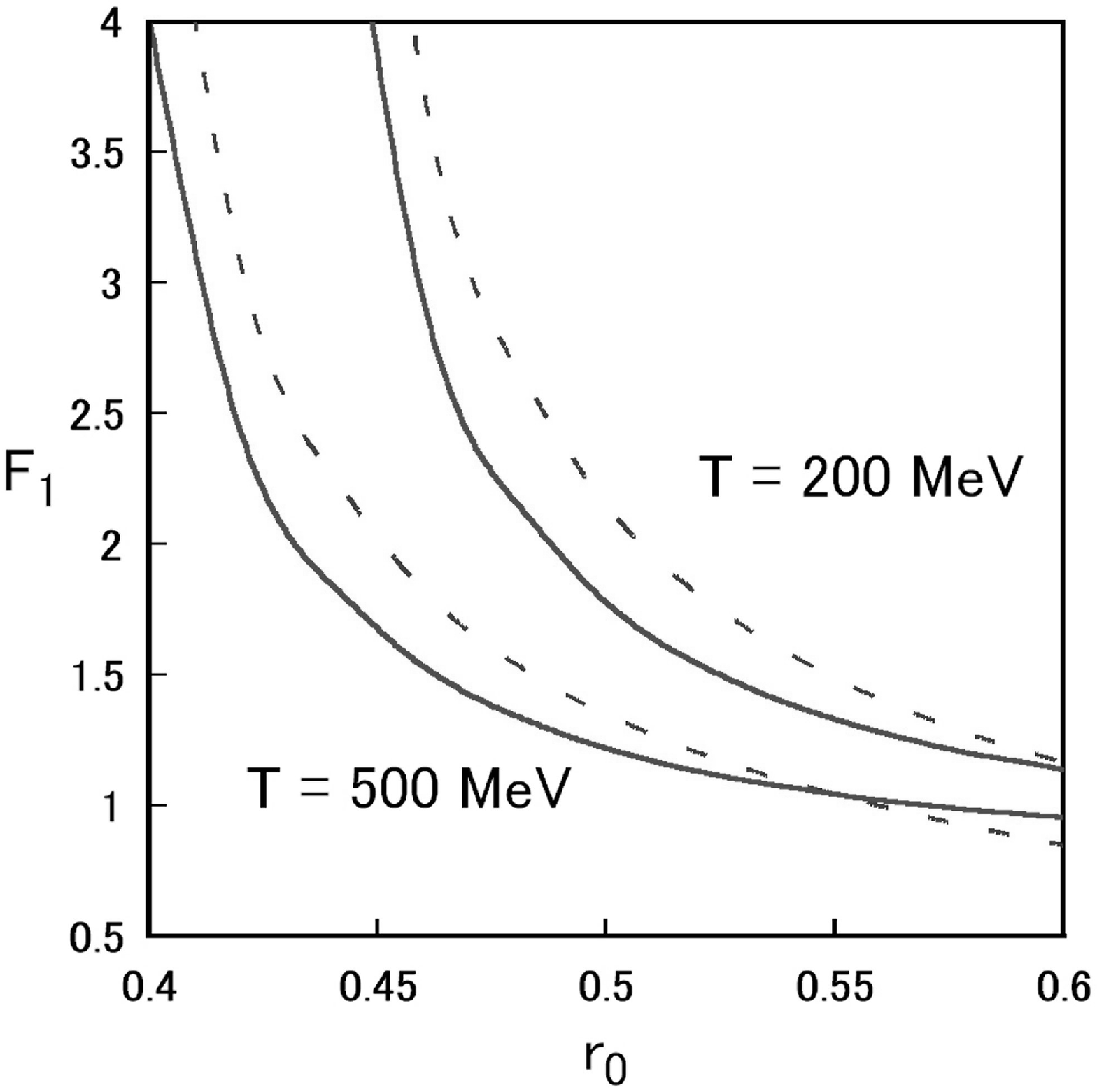}}
\caption{
The factor $F_1$ as a function of the core radius $r_0$ in units of fm.
The solid line shows $F_1$ with the Neumann function part as given in Eq. (34).
The dashed line shows $F_1$ without the Neumann function part as given in Eq. (32).
The upper two curves are with the laboratory energy $T$ = 200 MeV of the incident nucleon.
The lower two curves are with the laboratory energy $T$ = 500 MeV of the incident nucleon.
}
\end{center}
\end{figure}
\newpage
\begin{figure}
\begin{center}
\scalebox{0.5}{\includegraphics{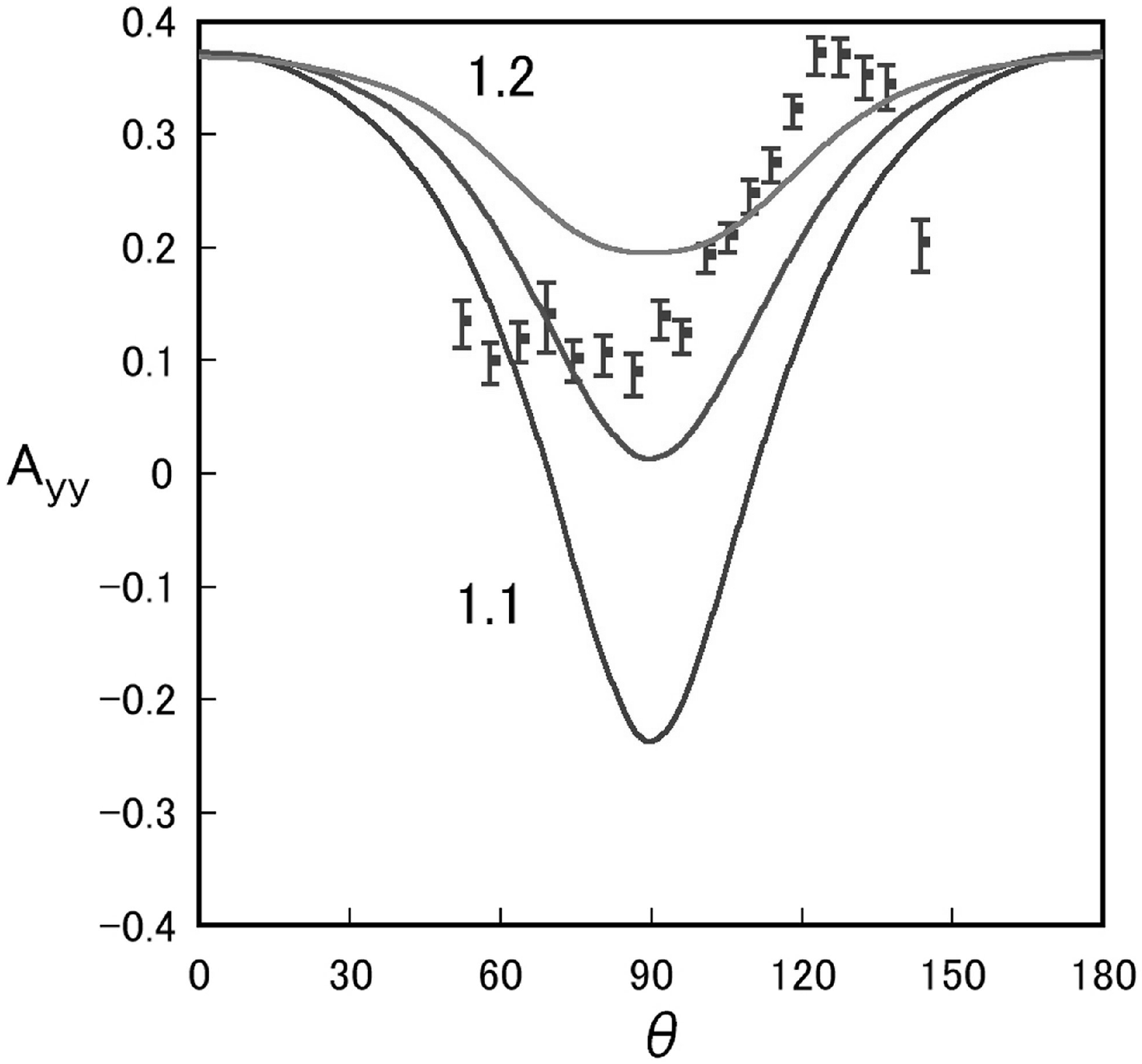}}
\caption{
The spin correlation parameter $A_{yy}(\theta)$ of neutron-proton system 
as a function of the center of mass scattering angle $\theta$ in degree at the laboratory energy $T$ = 425 MeV.
The three curves with $F_1$ = 1.1, $F_1$ = 1.15 and $F_1$ = 1.20 are shown in ascending order. 
The experimental data is from Ref. $\cite{Bandyopadhyay}$.
}
\end{center}
\end{figure}

\small
\begin{thebibliography}{99}
\bibitem{Seto}N. Set$\hat{\rm o}$, Prog. Theor. Phys. Suppl. {\bf 95}, 25(1988).
\bibitem{Kinpara}S. Kinpara, arXiv:nucl-th/1609.03294.
\bibitem{Case}K. M. Case, Phys. Rev. {\bf 80}(1950)797.
\bibitem{Bandyopadhyay}D. Bandyopadhyay, R. Abegg, M. Ahmad, J. Birchall, K. Chantziantoniou, C. A. Davis, N. E. Davison,
P. P. J. Delheij, P. W. Green, L. G. Greeniaus, D. C. Healey, C. Lapointe, W. J. McDonald, C. A. Miller, G. A. Moss,
S. A. Page, W. D. Ramsay, N. L. Rodning, G. Roy, W. T. H. van Oers, G. D. Wait, J. W. Watson and Y. Ye,
Phys. Rev. {\bf C40}(1989)2684.
\end{thebibliography}
\end{document}